\documentclass[aps,pra,twocolumn,nopacs]{revtex4-1}
\usepackage{graphicx}
\usepackage{amssymb,amsfonts,amsmath}
\usepackage[colorlinks=true,citecolor=Cerulean,linkcolor=RubineRed,urlcolor=Cerulean]{hyperref}
\hypersetup{breaklinks=true}
\usepackage{graphicx}
\usepackage{color}
\usepackage[usenames,dvipsnames]{xcolor}
\usepackage{epstopdf}
\usepackage{units}

\newcommand{\Heff}{H_\mathrm{eff}}
\newcommand{\Jeff}{J_\mathrm{eff}}
\newcommand{\Hlatt}{H_\mathrm{latt}}

\begin{document}

\title{Generating soliton trains through Floquet engineering}

\author{P.~Blanco-Mas}
\affiliation{Departamento de F\'isica de Materiales, Universidad
Complutense de Madrid, E-28040 Madrid, Spain}

\author{C.E.~Creffield}
\affiliation{Departamento de F\'isica de Materiales, Universidad
Complutense de Madrid, E-28040 Madrid, Spain}

\date{\today}

\begin{abstract}
We study a gas of interacting ultracold bosons held in a parabolic trap
in the presence of an optical lattice potential. Treating the system
as a discretised Gross-Pitaevskii model, we show how 
Floquet engineering, by rapidly ``shaking'' the lattice, 
allows the ground-state of the system to be 
converted into a train of bright solitons by inverting the sign of the hopping energy.
We study how the number of solitons 
produced depends on the system's nonlinearity and the curvature of the 
trap, show how the technique can be
applied both in the high and low driving-frequency regimes, and
demonstrate the phenomenon's stability against noise.
We conclude that the Floquet
approach is a useful and stable method of preparing solitons in
cold atom systems.
\end{abstract}

\maketitle

\section{Introduction}
A soliton is a localized excitation in a medium that preserves its
shape in time as it travels. First observed in  water waves
propagating in narrow channels \cite{first_soliton}, they are
ubiquitous in Nature, arising in such diverse contexts as
laser pulses in optical fibers \cite{laser_pulses}, the dynamics of tsunamis \cite{ocean}, and 
kinks moving along DNA \cite{dna}. Their stability arises from a balance between
a localized wavepacket's intrinsic tendency to spread in time,
and a non-linear attractive interaction which opposes this
spreading.

Bose-Einstein condensates (BECs) are particularly interesting
candidates to study soliton formation, as the 
systems are very clean and highly controllable,
and, in particular, the interatomic interaction can be manipulated accurately in the laboratory.
When the interaction is repulsive the BEC is stable. 
However, if the interaction is rapidly changed to be attractive
the BEC is rendered vulnerable to modulational instability, causing it to break up into
ripples which are then focused by the interaction to form solitons.
In this way,
soliton trains have been produced in BECs of
lithium-7 atoms \cite{strecker,khayovich}, rubidium-85 \cite{cornish}
and in cesium-133 \cite{caesium_1,caesium_2},
using the Feshbach resonance technique
to rapidly change the interaction strength. 
More recently, an alternative method of controlling the
atomic interaction has been developed which involves coupling internal 
atomic states with different scattering lengths with an RF field, which
has been successfully used \cite{leticia} to produce solitons in BECs of potassium-39.

In this work we revisit a protocol proposed by Carr and Brand (CB)
\cite{carr_brand_pra,carr_brand_prl}
for converting a trapped BEC into a soliton train. 
It consists of two parts. Initially a BEC with a repulsive interaction is held in
a trapping potential, as normal. This stable situation is then perturbed by
flipping the sign of the interatomic interaction, while simultaneously inverting
the trap so that it becomes expulsive.
If we write the Hamiltonian of the system as a sum of the kinetic energy $T$, the
trapping potential $V$, and the interaction term $U$, we can schematically
represent this process as
\begin{equation}
H = T + V + U \ \Longrightarrow T - V - U \ .
\end{equation}
This protocol thus requires control over both the interaction strength and the
sign of the trap, which may not always be experimentally feasible.
We instead propose to use a variant of this technique by addressing
a {\em single} parameter: the sign of the kinetic energy. This process can be
represented instead as
\begin{equation}
H = T + V + U \ \Longrightarrow -T + V + U \ ,
\label{scheme}
\end{equation}
which is clearly equivalent to the CB protocol, but with an overall minus sign.

The required inversion of the kinetic energy, equivalent to endowing the atoms with
a negative effective mass, can be achieved by a technique known as ``Floquet
engineering'' \cite{eckardt} by applying an external driving field which
oscillates periodically in time. The dynamics of the system can then be factored
into a rapid micromotion oscillating at the same frequency as the driving
field, and an effective static Hamiltonian $\Heff$. The parameters of $\Heff$
can be controlled very precisely by the driving field, and in particular
``shaking'' a tight-binding lattice model by rapidly oscillating the lattice in
space, allows the intersite tunneling to be coherently manipulated. The intersite
tunneling, for example, can be tuned to zero to produce the effect known
as ``coherent destruction of tunneling'' (CDT) \cite{hanggi}, it can be rendered complex
to allow the system to simulate the effect of a synthetic magnetic field 
\cite{synthetic_1,synthetic_2,flux}, or its sign can be inverted \cite{lignier,arimondo}, 
to provide a negative effective mass. 

Inverting the sign of the effective mass has been previously used in a static system to produce
gap solitons \cite{oberthaler}, by moving a trapped condensate to the edge of
the first Brillouin zone (FBZ) where the dispersion relation has negative curvature. More recently, Ref. \cite{haller}
performed an experiment
using Floquet engineering to flip the sign of the tunneling by high-frequency shaking
of a cesium BEC
to obtain a solitonic wavepacket at the center of the FBZ.
By employing a similar driving of this type we will show how
Floquet engineering can be used in conjunction
with varying the trapping potential and the magnitude of 
the atomic interaction to create stable soliton trains containing
a specific number of solitons. We will
then go on to examine both the high and low frequency driving regimes, and show that
soliton formation occurs in both. Having the ability to vary the driving frequency in
this way gives the flexibility to avoid exciting the atoms to a higher band
at high driving frequencies, and also to evade parametric resonances \cite{bukov}, 
which would otherwise heat and eventually destroy
the condensate. Finally we will study the effect of the phase of the driving on the
protocol, and demonstrate the scheme's remarkably high robustness to noise.

\section{Method}

\subsection{Model}
A BEC held in a trap potential can be described well by the Gross-Pitaevskii Hamiltonian
\begin{equation}
H_{\mathrm GP} = \frac{- \hbar^2}{2 m} \partial_x^2 + V(x) + g \left| \psi(x) \right|^2 \ ,
\label{gpe}
\end{equation}
where $g$ is the interatomic coupling constant, proportional to the $s$-wave scattering length, 
$m$ is the atomic mass, and the condensate wavefunction, $\psi(x)$, is normalized to one.
The trapping potential is denoted by
$V(x)$, and will be taken to be quadratic, $V(x) = V_0 x^2$, where the distance $x$ is
measured from the center of the system.
If we now discretise space by imposing an optical lattice potential, Eq. \ref{gpe} can
be rewritten in second-quantized form,
\begin{equation}
\Hlatt = -J \sum_j \left( a_j a_{j+1}^\dagger + \mathrm{H.c.} \right) +
\sum_j V(x_j) n_j + g \sum_j n_j^2 \ ,
\label{discrete}
\end{equation}
where the kinetic energy is now given in terms of the tunneling $J$ between nearest-neighbor
lattice sites, $a_j / a_j^\dagger$ are the annihilation / creation operators for
a boson on site $j$,
and $n_j = a_j^\dagger a_j$ is the standard number operator.
Henceforth we will take $\hbar = 1$, and measure all energies and
frequencies in units of $J$, and write all distances in units of the lattice constant.

\begin{figure}
\begin{center}
\includegraphics[width=0.5\textwidth,clip=true]{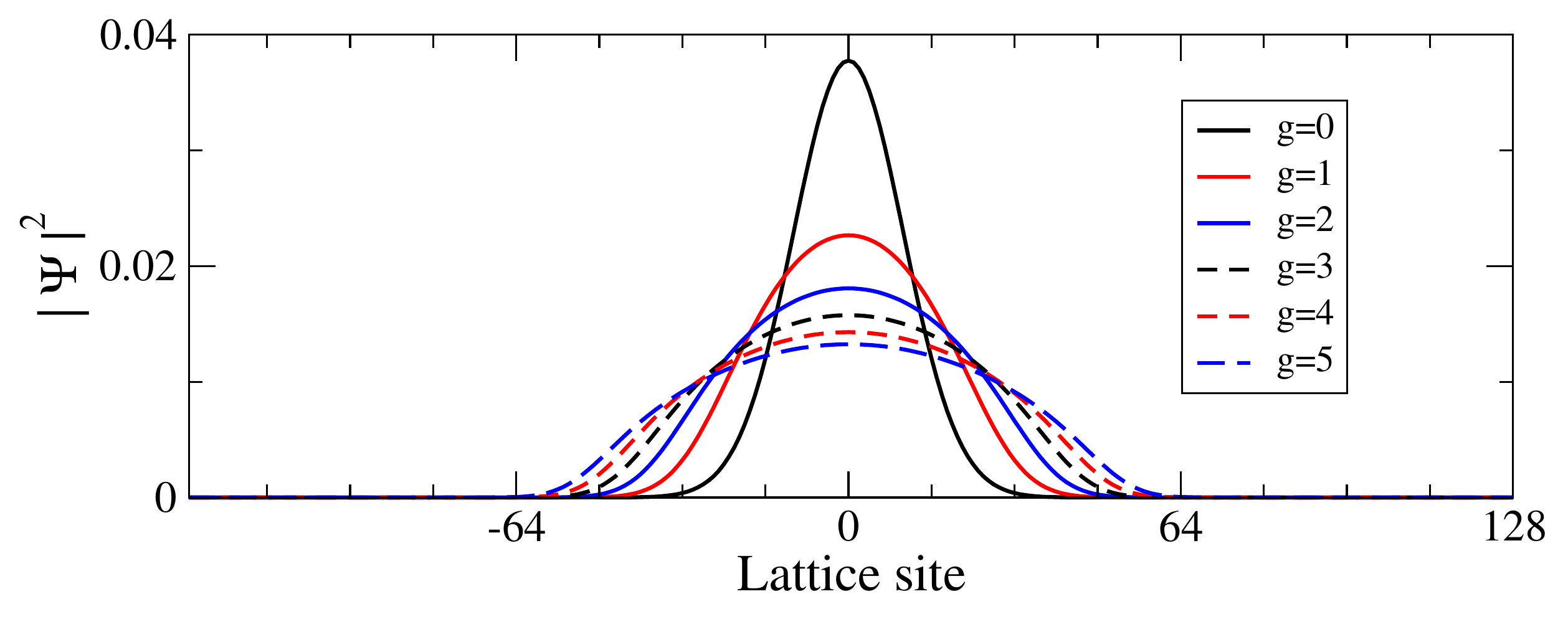}
\end{center}
\caption{Probability density of the ground state of the
lattice GPE (Eq. \ref{discrete}) for different values of the non-linearity $g$.
The curvature of the trap is given by $V_0 = 2 \times 10^{-5}$.
Note how the wavefunction evolves from a
Gaussian for $g=0$, to a broader, more flattened form as $g$
increases and the system approaches the Thomas-Fermi limit.}
\label{initial_state}
\end{figure}

We will take the initial state for the simulations to be the ground state of Hamiltonian 
(\ref{discrete}). This is obtained by starting with the solution for $g=0$ 
(a Gaussian) and evolving it under Eq. \ref{discrete}
in imaginary time, maintaining the correct normalization, until convergence is achieved. 
The results for several different values of $g$ are shown in Fig. \ref{initial_state}, 
and the evolution from a narrow Gaussian
distribution to the broader inverted parabolic form predicted by the Thomas-Fermi approximation
in the limit of large $g$ is clearly visible.

We now wish to introduce the time-dependent driving potential. If the optical lattice
is periodically oscillated in space, or ``shaken'', an observer in the rest frame
of the lattice perceives an inertial force described by a time-dependent
lattice tilt, resulting in the Hamiltonian
\begin{equation}
H(t) =  \Hlatt + K(t) \sum_j j \ n_j \ ,
\label{shaking}
\end{equation}
where $K(t)$ describes the form of the shaking. For the common case of sinusoidal shaking,
this function can be written as
$K(t) = K \cos \left( \omega t + \phi \right)$, where $\omega$ is the frequency
of the driving, $K$ is its amplitude,
and for generality we have included a driving-phase $\phi$.
All numerical results were obtained by first preparing the initial
state using the imaginary time relaxation method described above,
and then integrating this state in time under Eq. \ref{shaking},
using a fourth-order Runge-Kutta routine.

\subsection{Floquet engineering}
As the time-dependent driving (\ref{shaking}) is $T$-periodic, $H(t) = H(t + n T)$, 
solutions of the time-dependent Schr\"odinger equation are of Floquet form
\begin{equation}
\left[ i \partial_t - H(t) \right] \psi_n(t) = \epsilon_n \psi_n(t)
\label{floquet}
\end{equation}
where the Floquet states $\psi_n(t)$ have the same $T$-periodicity as the Hamiltonian,
and provide a complete basis to describe the time evolution of the driven system. 
The eigenvalues $\epsilon_n$ are called the quasienergies, and govern the long-term
dynamics of the system, as the time-dependence of the Floquet states only operates over 
timescales within each driving period, providing the so-called ``micromotion'' \cite{eckardt}
of the system.

\begin{figure}
\begin{center}
\includegraphics[width=0.5\textwidth,clip=true]{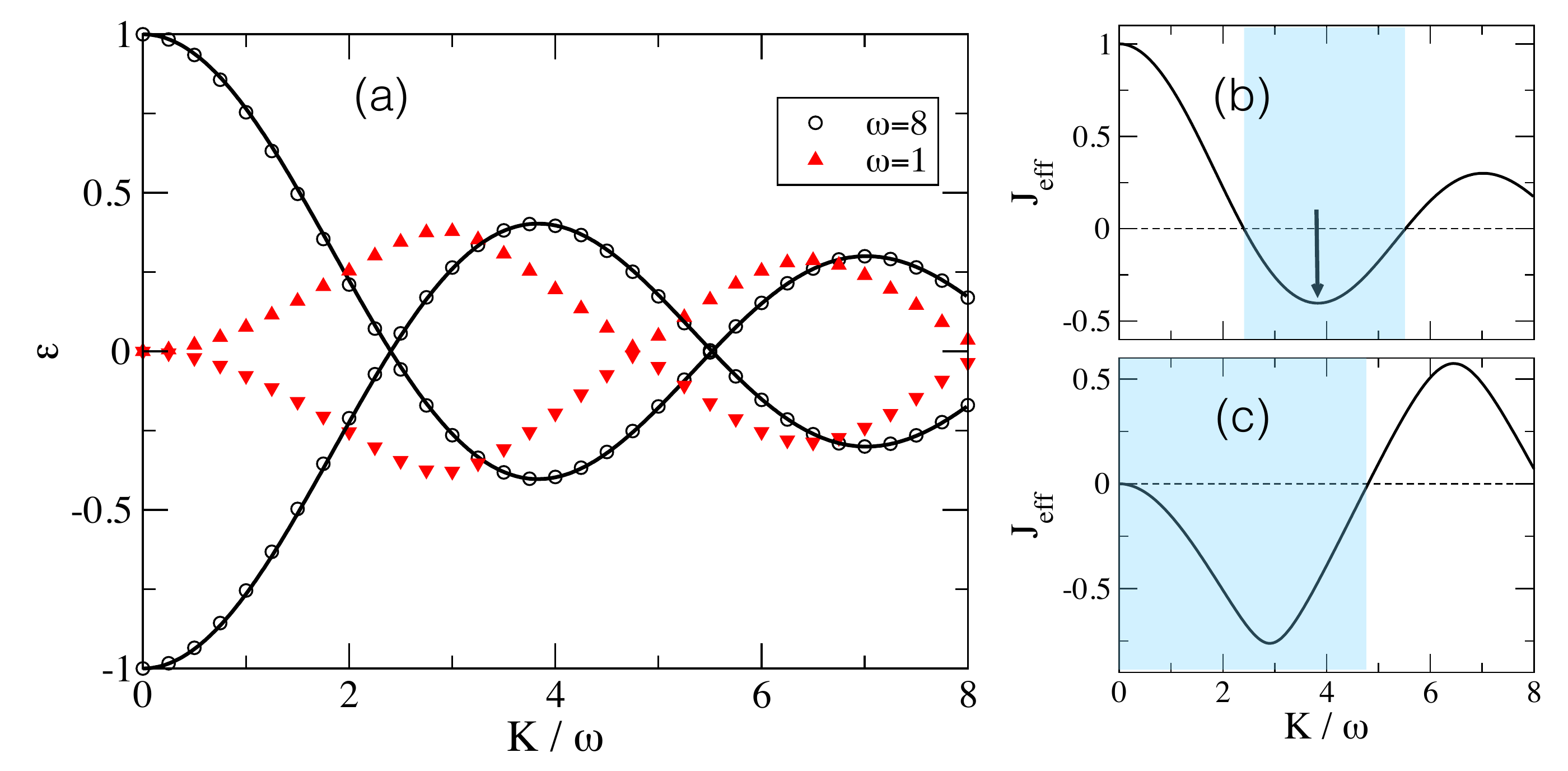}
\end{center}
\caption{(a) Quasienergies for the two-level model. Black solid lines
show the first-order approximation $\epsilon = \pm {\cal J}_0 \left(K / \omega \right)$ 
which becomes exact in the limit $\omega \rightarrow \infty$.
Black circles show the exact numerical results for $\omega = 16$; at this high-frequency
the results are excellently approximated by the perturbative result.
The red triangles show the quasienergies for $\omega = 1$, which differ considerably
from the Bessel function behaviour of the high frequency results.
b) $\Jeff$ for $\omega = 16$;
note how this vanishes at $K / \omega =2.404$, and is negative in the
shaded region between the first and second zeros of ${\cal J}_0$.
The arrow marks $K / \omega = 3.80$, the value used in the Floquet engineering method.
c) $\Jeff$ for $\omega = 1$. Again, the shaded region indicates where 
the effective tunneling is negative. The quantities
$\epsilon$ and $\Jeff$ are measured in units of $J$
(see text).}
\label{quasienergies}
\end{figure}

The quasienergies can be obtained as the eigenenergies of an effective static
Hamiltonian, $\Heff$, which depends on the parameters and form of the driving.
The process of Floquet engineering then consists of choosing the appropriate driving function
to produce the desired form of $\Heff$, and thus the quasienergies. 
In general it is difficult to obtain closed-form solutions for the quasienergies for
a given drive. It is possible, however, to obtain expressions for $\Heff$ as series
expansions in inverse frequency, such as the Magnus series \cite{magnus}
and the Van Vleck series \cite{van_vleck}, 
which become exact in the limit of infinite driving frequency. To first order,
it can be shown that for a sinusoidally-driven two-level system the quasienergies
are given by \cite{crossings}
$\epsilon_\pm = \pm J \ {\cal J}_0 \left( K / \omega \right)$,
where ${\cal J}_0$ is the zeroth Bessel function of the first kind. 
Fig. \ref{quasienergies} shows the excellent agreement between this result
and the numerical results for a two-level system driven at $\omega = 16$. Good agreement
continues to be obtained as long as $\omega > J$: the high-frequency regime.
However, for lower frequencies the quasienergies behave differently, 
as can be seen for the case of $\omega = 1$, indicating that
more terms \cite{barata} must be included in the series. 

The effective tunneling between the levels is proportional to the difference
between the quasienergies, $\Jeff = \left( \epsilon_+ - \epsilon_- \right)/2$,
and thus in the high-frequency limit the effective tunneling is given by
\begin{equation}
\Jeff =  J \ {\cal J}_0 \left( K / \omega \right)  \ .
\label{bessel}
\end{equation}
Altering the parameter $K / \omega$ therefore allows $\Jeff$ to be tuned to a
desired value. In particular, if we set $K / \omega =2.404$ (the first zero of
the Bessel function) the two quasienergies become degenerate,
as can be seen in Fig. \ref{quasienergies}a, and the effective
tunneling vanishes, producing CDT. Increasing $K / \omega$ beyond this value causes
$\Jeff$ to become negative in the interval between the first and second zeros
of ${\cal J}_0$ (the shaded region in Fig. \ref{quasienergies} b), 
which is thus the region of interest for our soliton
generation method (\ref{scheme}).

Proceeding to an $N$-site lattice system, the quasienergies now present a band-like structure 
\cite{instability}
\begin{equation}
\epsilon_n = -2 \Jeff \cos k_n  \ ,
\label{band}
\end{equation}
where $k_n$ are the permitted momenta in 
the FBZ, and $\Jeff$ is the same intersite tunneling discussed above (Eq. \ref{bessel}). 
Accordingly we can
regulate the effective mass of a particle moving in the lattice in the same way that we can
control the tunneling in the two-level model, by appropriately choosing the
value of $K / \omega$.

\section{Results}

\begin{figure*}
\begin{center}
\includegraphics[width=1.0\textwidth,clip=true]{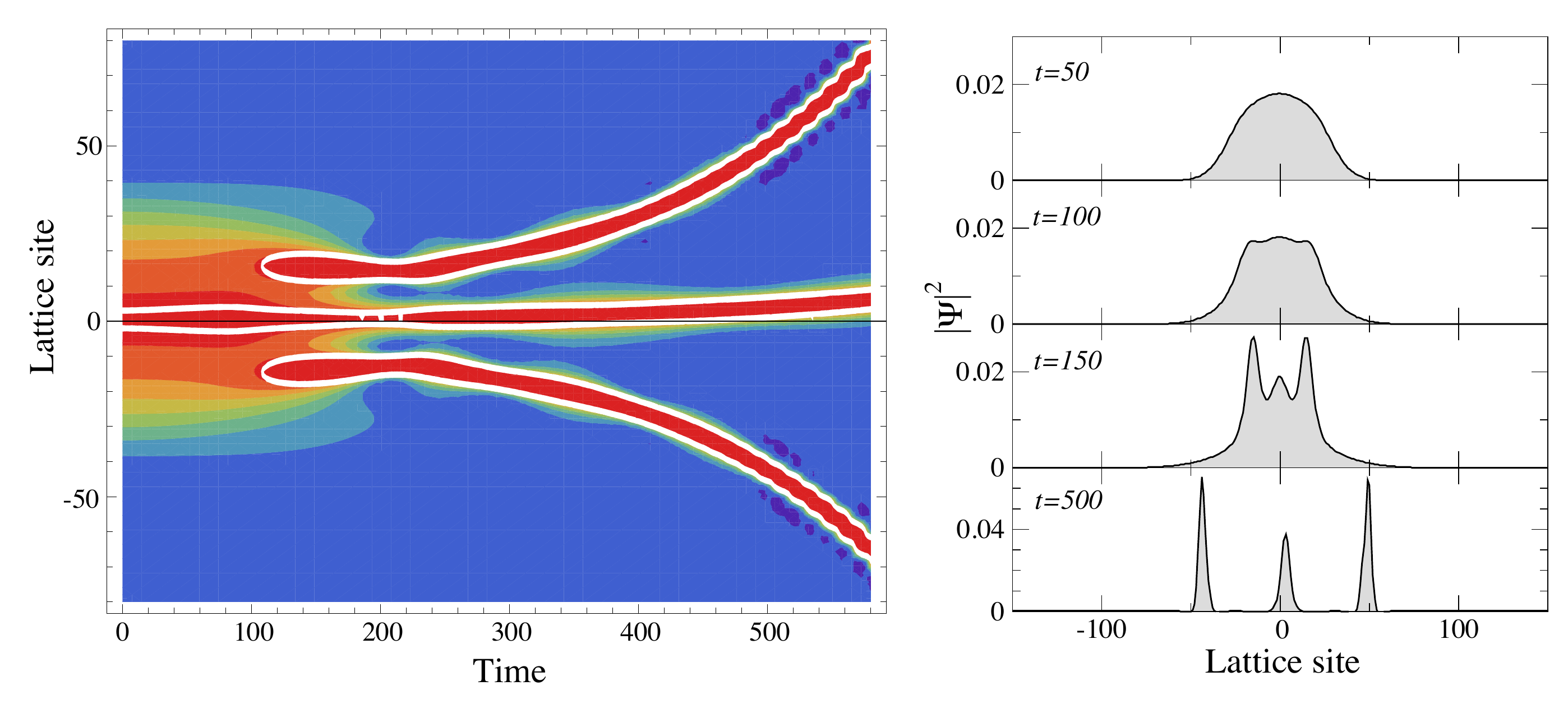}
\end{center}
\caption{Soliton generation using the Floquet engineering protocol (\ref{scheme}).
Physical parameters: 512 lattices sites, $V_0 = 2 \times 10^{-5}$, $g=2$, and
the driving frequency $\omega = 16$.
Left: contour plot of the condensate density $|\psi(x,t)|^2$. The system is initialized
in its ground state, and at $t=0$ the driving potential is applied. At $t=100$
ripples begin to form in the condensate, which gradually focus into three
solitons. The central soliton remains close to the center of the lattice, remaining
in unstable equilibrium with the lattice potential, while the other two solitons
accelerate exponentially away.
Right: cross sections through the condensate at different times. The ripples form
initially at the edges of the wavepacket, indicating they arise from self-interference,
and then sharpen into the typical soliton shape.} 
\label{solitons}
\end{figure*}

{\em Soliton generation -- }
In Fig. \ref{solitons} we show the results of applying the Floquet protocol to
a system with nonlinearity parameter $g=2$.
The system was prepared in its ground state, and then at $t=0$ the driving
potential was suddenly turned on, with phase $\phi = 0$, and a frequency of $\omega = 16$
placing the system firmly in the high-frequency regime.
As shown in Fig. \ref{quasienergies}b, the amplitude of the driving was chosen to be 
$K / \omega = 3.80$ 
that is, the first minimum of Eq. \ref{bessel}, the corresponding
value of $\Jeff = -0.403$ being the largest negative value possible for
the effective tunneling.

When the driving potential is applied, small ripples develop in the
profile of the condensate. As Carr and Brand demonstrated, the
dominant source of these ripples is self-interference of the condensate
wavefunction, which causes them to develop first at the edges of the wavepacket.
Modulational instability causes the ripples to grow
in amplitude, and they are then focused by the nonlinearity into forming sharp peaks with
a typical soliton profile.
For this value of the nonlinearity the initial wavepacket divides into 
three solitons. The one which forms near the center of the trap remains
essentially stationary over the remainder of the time evolution, while
the other two accelerate away at an exponentially increasing velocity.
Although the trapping potential does not change during the protocol, the solitons 
have a negative effective mass due to the inversion of $\Jeff$, and so ``fall uphill''
against the potential and are accelerated outwards, just like a normal
particle in an inverted potential. Note that once formed, the solitons are stable,
and retain their form throughout their trajectory.

\begin{figure}
\begin{center}
\includegraphics[width=0.5\textwidth,clip=true]{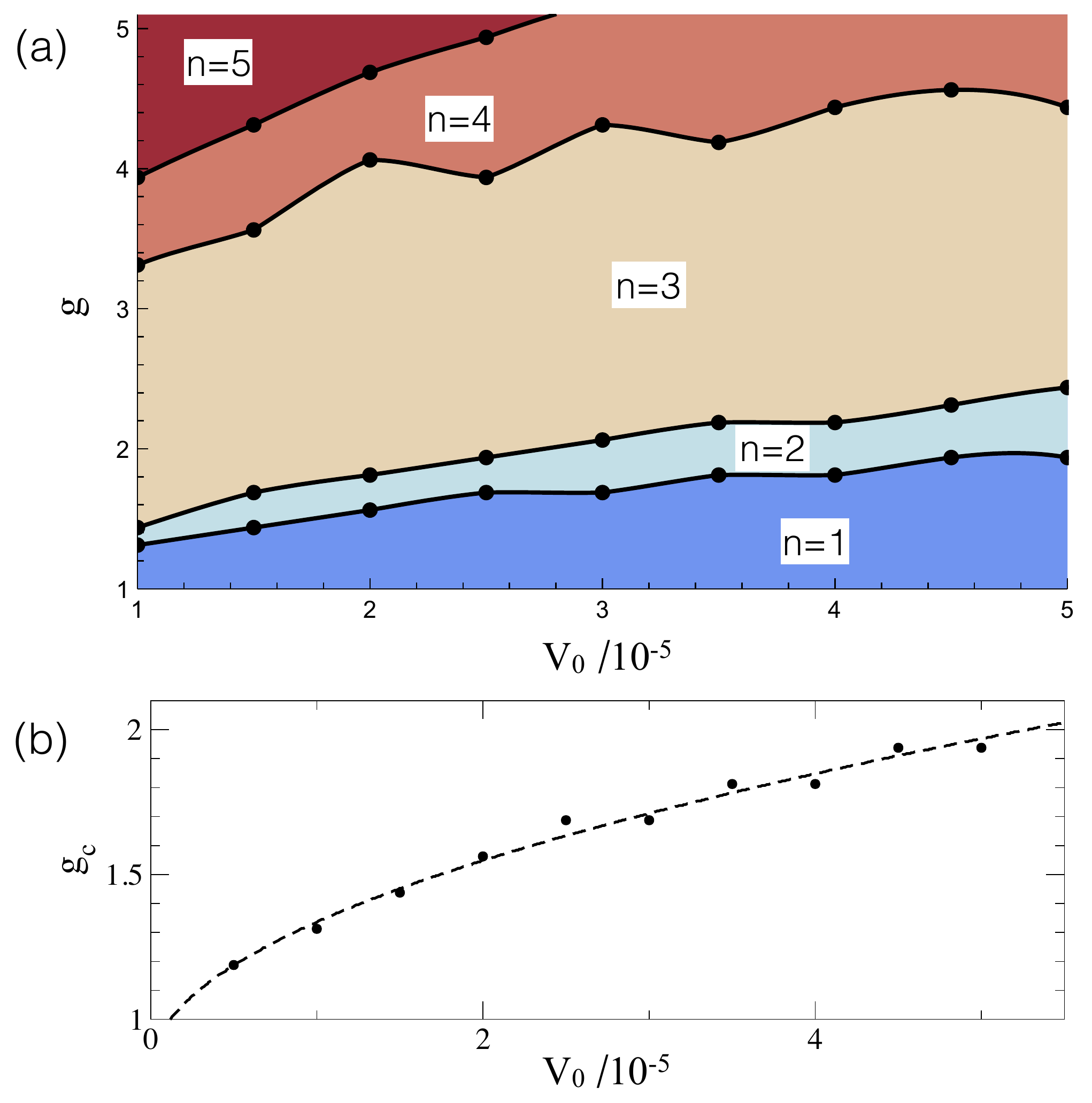}
\end{center}
\caption{a) Number of solitons produced as the trap curvature $V_0$ and
the nonlinearity $g$, both measured in units of $J$, are varied. 
The number of solitons increases for
high $g$ and low $V_0$, as predicted by Eq. \ref{scaling}.
In all cases a lattice of 256 sites was used.
b) Plot of the critical interaction strength, $g_c$, at which the system
changes from producing two solitons to producing three. The dashed line is
a least squares fit to $g_c = a_1 + a_2 \sqrt{V_0}$, where $a_1$ and
$a_2$ are fitting parameters, which describes
the behaviour with reasonable accuracy.}
\label{phase_diagram}
\end{figure}

Raising the value of $g$ reveals that
the number of solitons produced in a trap with a given curvature increases
weakly as a function of the nonlinearity,
as observed previously in Refs. \cite{hulet,everitt}. 
The trap curvature  can also be used as a parameter to control the soliton
number.  We show the combined effect of these factors in Fig. \ref{phase_diagram}a.
From this plot it is clear that high soliton numbers are favored by
a high value of the nonlinearity and a low trap curvature. This can be
understood qualitatively by a simple scaling argument.
The growth of the ripples
is governed by the most unstable Bogoliubov mode, the wavelength of which \cite{hulet}
is related to the healing length of the condensate, $\xi$.
The other length scale of the problem is the harmonic oscillator length $\ell$
which gives an estimate for the effective width of the initial wavepacket.
On dimensional grounds, the number of solitons
will vary approximately as the ratio of these lengths
\begin{equation}
n \sim \xi / \ell \sim \left( g / \sqrt{V_0} \right)^{1/2} \ , 
\label{scaling}
\end{equation}
in agreement with the trends observed.

To examine this behaviour more quantitatively, we plot in Fig. \ref{phase_diagram}b the critical
value of the nonlinearity parameter, $g_c$, at which the number of solitons
produced in the system changes from two to three -- that is, the lowest boundary
curve plotted in Fig. \ref{phase_diagram}a. According to Eq. \ref{scaling},
this quantity should vary as $g_c \sim \sqrt{V_0}$. As can be seen, the data can
indeed be fitted with reasonable accuracy by this expression.

\medskip
{\em Low-frequency regime -- }
The results presented so far have been for $\omega = 16$, which is well within the
high-frequency regime. The results obtained are essentially identical to those
obtained by performing the simulations without the time-dependent driving, but setting
the value of $J$ to $\Jeff = -0.403$ by hand at $t = 0$, which we shall term
the ``switched protocol''. The excellent agreement between the results
indicates how well the Floquet protocol duplicates the switched protocol
for this frequency. 

As $\omega$ is lowered we should expect this agreement to reduce, as more
terms must be included in the Magnus expansion,
and $\Heff$ can no longer be approximated as a nearest-neighbor hopping Hamiltonian
with a single effective tunneling.
Perhaps surprisingly, although
the amplitude and position of the generated solitons alter slightly for
smaller values of $\omega$, the process of soliton formation itself remains robust.
The important point is just that the nearest-neighbor tunneling must become
negative. Although the positions of quasienergy degeneracies will drift
away from the zeros of the Bessel function as the frequency is reduced
\cite{crossings}, there will nonetheless still be some intervals over which $\Jeff$ will
be negative. 

As an example, in Fig. \ref{quasienergies}c, we plot the
effective tunneling for a system at a low driving frequency of $\omega = 1$.
As we noted previously, $\Jeff$ is related to the difference of the
quasienergies, but unlike the high-frequency case, in this instance
we do not know which quasienergy corresponds to $\epsilon_+$ and
which to $\epsilon_-$. As a result, even knowing the values of the quasienergies
we are uncertain of the sign of $\Jeff$. Simulating the driven system
in a lattice with an additional static tilt reveals that an initial wavepacket moves
up the potential for $K / \omega < 4.8$, indicating that its effective
mass is negative, while it moves down the potential for higher driving
amplitudes, corresponding to its effective mass being positive.
Accordingly, driving the system at $K / \omega = 3.8$, as done in the high-frequency
case, should also produce solitons.

\begin{figure}[t]
\begin{center}
\includegraphics[width=0.5\textwidth,clip=true]{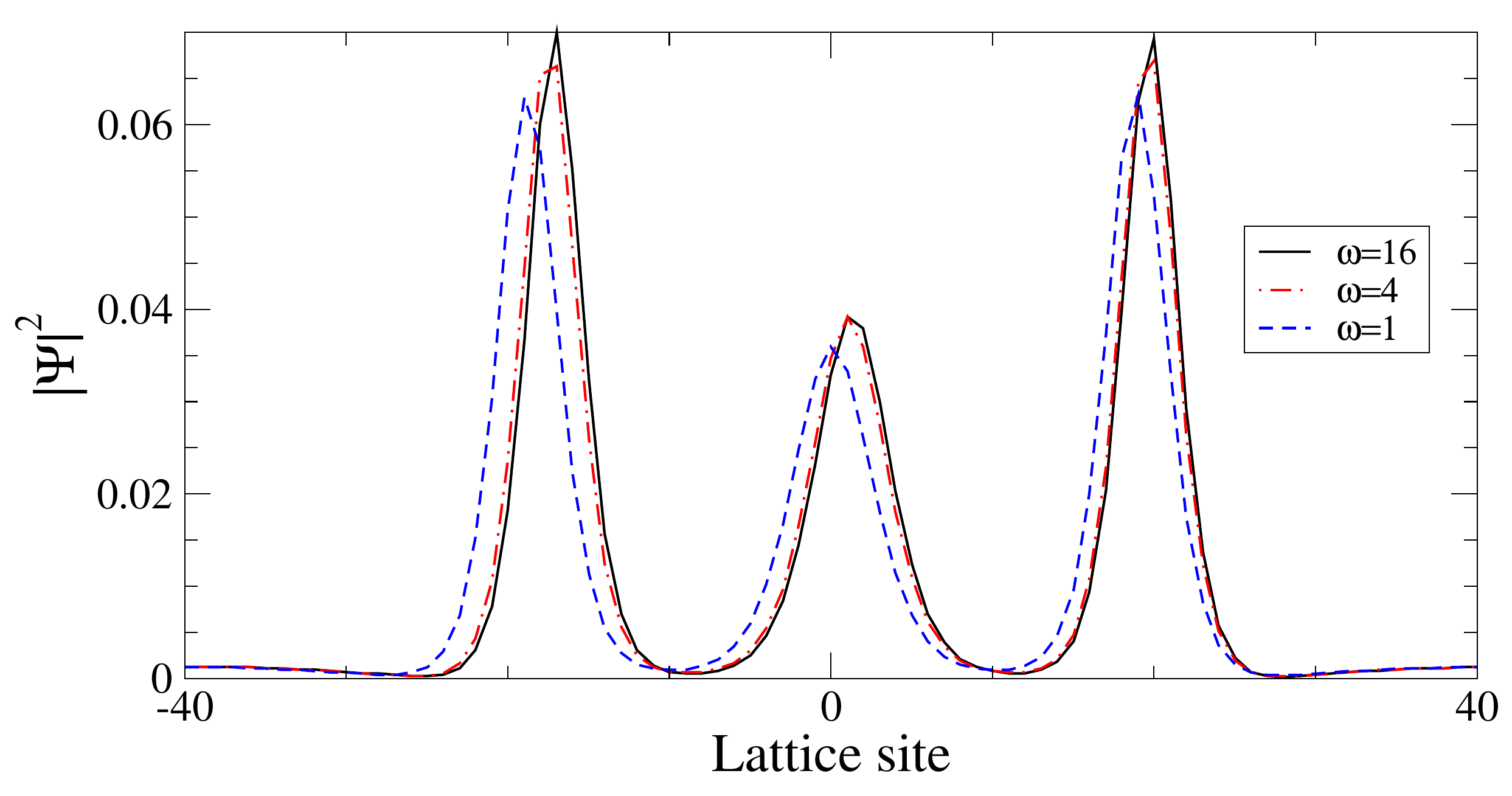}
\end{center}
\caption{Soliton development at $t=300$ for three different driving frequencies.
In the high frequency regime, $\omega = 16$ (solid black line), 
the results obtained are indistinguishable
from the switched protocol. For $\omega=4$ (dash-dotted red line) the solitons produced strongly resemble
the high frequency result, although small differences are visible. At low frequency,
$\omega = 1$ (dashed blue line), the differences are larger, but qualitatively a very similar result
is still obtained. System parameters: 512 lattice sites, $V_0 = 0.00002$ and $g=2$.}
\label{freq}
\end{figure}

We show the results in Fig. \ref{freq} for  three different driving frequencies. At
the sample time used, $t= 300$, the switched protocol produces in three solitons.
For a driving frequency of $\omega = 16$, the result is essentially
indistinguishable for the switched protocol. Lowering the driving frequency
to $\omega = 4$ again gives a very similar result: although the three solitons produced
have slightly different positions and amplitudes, the differences from the $\omega=16$
result are very minor. Taking now the value $\omega = 1$, well outside
the high-frequency regime, again gives a very similar result. Using a high
driving frequency is thus not a necessary requirement for this technique, as
long as $\Jeff$ changes sign.

\medskip
{\em Effect of phase -- }
The amplitude of the driving is set through the requirement $K / \omega = 3.8$,
and we have shown that the effect is robust to varying the driving frequency, $\omega$.
This leaves one last parameter in the driving to consider: the phase, $\phi$. We have so far
used cosinusoidal driving, that is, $\phi = 0$. If we instead use sinusoidal driving ($\phi = \pi/2$)
we see no sign of any soliton creation. The wavepacket instead sloshes from side to side
of the trapping potential. This effect is produced by the so-called kick-operator \cite{eckardt}.
As shown in Ref. \cite{phase}, when $\phi$ is not zero, the driving imprints a phase
onto the condensate which excites it into motion. For the type of driving  we consider
the full expression for the effective tunneling is given in the high-frequency limit by
\begin{equation}
\Jeff = J e^{-i \left[ K / \omega \right] \sin \phi} 
{\cal J}_0 \left( K / \omega \right) \ ,
\label{kick}
\end{equation}
which clearly reduces to Eq. \ref{bessel} for $\phi = 0$. While the same Bessel function
renormalization of the modulus of the tunneling still happens for sinusoidal driving, the additional
phase factors mean that $\Jeff$ does not flip sign when $K / \omega = 3.80$. The driving
instead delivers a kick to the condensate, giving it an initial velocity $v_0 = K / \omega$,
and thus causing it to make harmonic oscillations about the center of the trap.

\medskip
{\em Noise -- } 
We finally consider how stable the soliton creation process is to noise on the
initial state, since in experiment it is clearly impossible
to prepare the desired initial state with complete fidelity.
We have so far used the ground state, $\psi^0(x)$, as the initial state, and 
we will now add random noise to it
\begin{equation}
\psi(x,t=0) = \psi^0(x) + r(x) \gamma \ ,
\label{random}
\end{equation}
where $r$ is a random variable uniformly distributed over $(-1,1)$, and
$\gamma$ sets the amplitude of the noise. After the noise has been added to 
both the real and imaginary components of $\psi^0(x)$, the
resulting state is normalized to unity as usual, 
and used as the initial state of the simulation.

\begin{figure}
\begin{center}
\includegraphics[width=0.5\textwidth,clip=true]{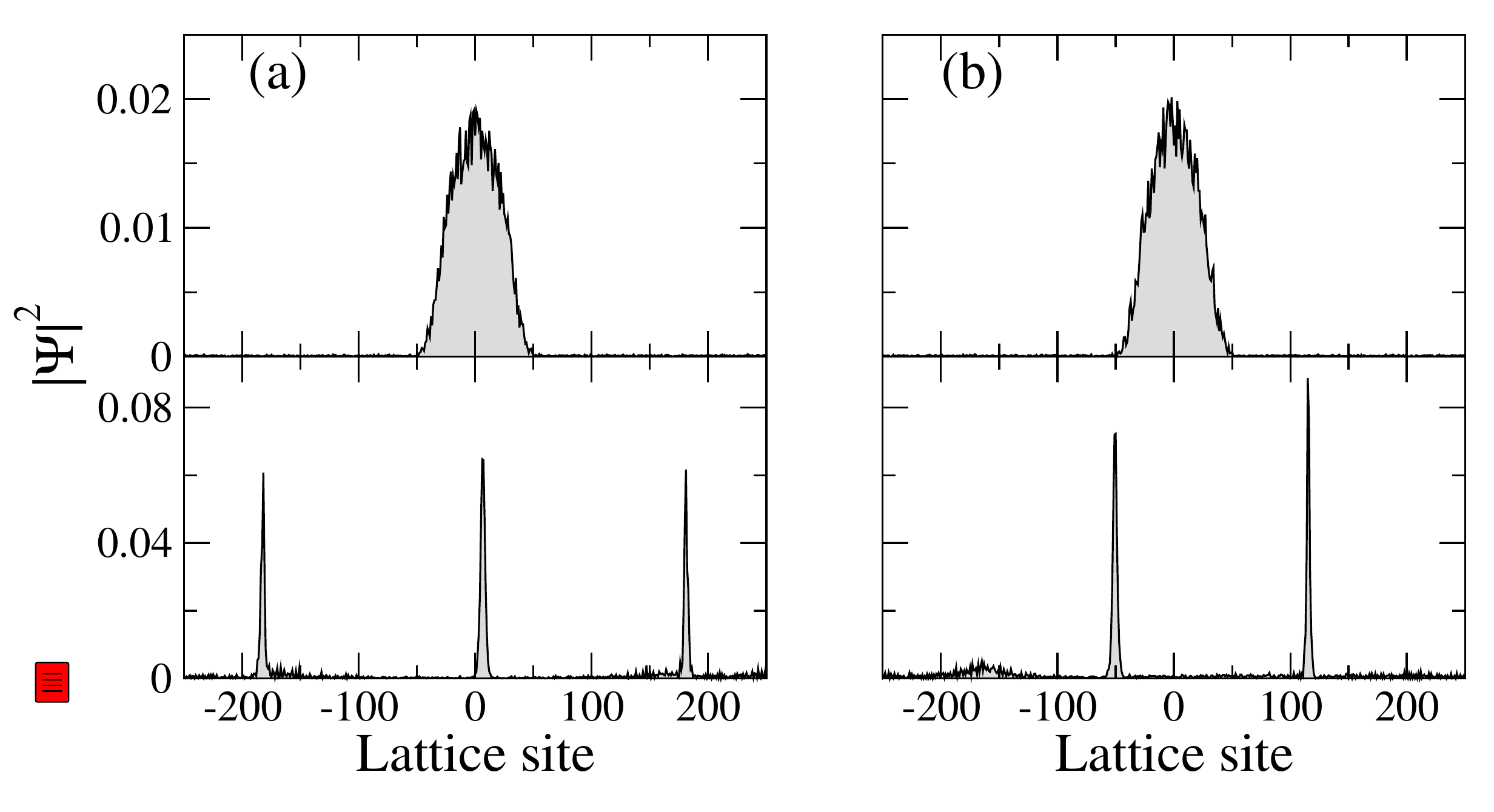}
\end{center}
\caption{Effect of noise on soliton generation.
(a) Above: initial state with $\gamma = 0.01$. Below: the particle
density of the system at $t=300$. Just as in the clean system
(Fig. \ref{solitons}), three solitons form.
(b) Above: initial state with $\gamma = 0.05$. Below: at
$t=300$ the system has evolved to present just two solitons,
the random noise having affected details of the soliton generation process.
Physical parameters: 512 lattice sites, $V = 2 \times 10^{-5}$, $g= 2$.}
\label{noise}
\end{figure}

We show the results for two different noise levels in Fig. \ref{noise}.
For the given system parameters, $V=2 \times 10^{-5}$ and $g = 2$,
we expect to produce three solitons,
as seen previously in Fig. \ref{solitons}. For $\gamma=0.01$ (Fig. \ref{noise}a) 
the initial state already appears notably irregular, but the process indeed gives
rise to three well defined solitons. 
Unlike in the clean system, the seeds for soliton formation are
now dominated by the imposed random fluctuations in the condensate wavefunction,
rather than by self-interference of the condensate. As a result
the positions of the solitons are randomly shifted in position with respect
to the clean system, and the solitons are more equal in size, as they began forming 
at essentially the same time. Further
increasing the noise level to $\gamma = 0.05$ produces two solitons instead of the
three expected. Nonetheless it is striking that even in the presence of such a
high level of noise, the process still produces clearly identifiable solitons that 
are easily detectable above the background noise level 

\section{Conclusions}
In summary, we have shown how Floquet engineering may be used to implement a
protocol based on the CB method for converting a trapped condensate into a
train of solitons. It has the advantage that only one control parameter needs
to be altered, and as it does not involve controlling the interparticle interaction,
it is applicable to atomic species for which this control is not easily available,
such as those like rubidium-87 which lack a convenient Feshbach resonance. 
The method consists
of using Floquet physics to invert the sign of the intersite tunneling,
or equivalently, to give the condensate atoms a negative effective mass.
Modulational instability then causes the initial state to break up into a
train of spatially-localized pulses, which then self-focus into solitons
under the influence of the nonlinear interaction.
We have demonstrated how the curvature of the trap and the magnitude of the
atomic interaction can be used to deterministically prepare trains of a
given number of solitons. Unlike many applications of Floquet engineering,
the method is not restricted to the high-frequency regime, but works equally
well for lower driving frequencies, which have the advantage of avoiding
driving the atoms into higher bands. In experiment, soliton generation in 
the low-frequency regime may be harder to attain due to the
increased rate of heating \cite{bukov} destroying the coherence of the condensate.
Nonetheless, experiments \cite{lignier} have reported coherent control of the tunneling
down to frequencies of $\omega = 0.3 J$, well into the low frequency regime,
and once formed the solitons are
able to self-cool \cite{carr_brand_pra} by emitting small bursts of atoms.
Finally we would like to emphasise the
remarkable stability of the method with respect to noise. The modulational
instability is seeded by the most unstable Bogoliubov mode, and as this typically has
a fairly long wavelength, noise on shorter length scales has relatively
little effect. The repeatability and excellent control afforded by this method
make it an excellent tool to investigate soliton dynamics and collisions, and
hold out the prospect of using such soliton trains for precision measurement applications
such as atom interferometry.

\acknowledgments
This work was supported by the Universidad Complutense de Madrid
through grant no. FEI-EU-19-12.

\bibliographystyle{aipnum4-1}
\bibliography{soliton_bib}

\end{document}